# Exactly solvable toy model for surface plasmon amplification by stimulated emission of radiation

D. G. Baranov,[1,2] E.S. Andrianov,[1,2] A. P. Vinogradov,[1,2] and A. A. Lisyansky[3]

[1]*Moscow Institute of Physics and Technology, 9 Institutskiy per., Dolgoprudny 141701, Moscow Reg., Russia*
[2]*Institute for Theoretical and Applied Electromagnetics, 13 Izhorskaya, Moscow 125412, Russia*
*Department of Physics, Queens College of the City University of New York, Flushing, NY 11367, USA*

We propose an exactly solvable quasi-classical model for surface plasmon amplification by stimulated emission of radiation (spaser). The gain medium is described in terms of the nonlinear permittivity with negative losses. The model demonstrates the main features of a spaser: a self-oscillating state (spasing) arising without an external driving field if the pumping exceeds some threshold value, synchronization of a spaser by an external field within the Arnold tongue, and the possibility of compensating for Joule losses when the pumping is below threshold. Similar to the common laser, a transition to the spasing regime takes a form of the Hopf bifurcation.

## I. INTORDUCTION

The spaser (Surface Plasmon Amplification by Stimulated Emission of Radiation) was first suggested in Ref. 1 and experimentally realized in Ref. 2. The spaser is a quantum device aimed at enhancing the near field of surface plasmons (SPs) excited on a metal nanoparticle (NP) by a quantum system, e.g. a quantum dot (QD), with population inversion. The physical principle of spaser operation is similar to that of a laser. The role of photons confined to the Fabri-Perot resonator is played by SPs[1, 3-5]. The NP and the QD are placed near to each other. SPs excited on the NP, therefore, trigger stimulated transition at the QD which in turn excites more SPs. The main difference between spasers and lasers is that a spaser generates and amplifies the nonradiative plasmonic mode of a NP in contrast to the radiative field of a conventional laser. This SP amplification occurs due to radiationless energy transfer from the QD to the NP. This process originates from the dipole-dipole (or any other near field[6]) interaction between the QD and the plasmonic NP. This physical mechanism is highly efficient because the probability of SP excitation is approximately $(kL)^{-3}$ times larger[7] than the probability of radiative emission, where $L$ is the distance between the centers of the NP and the QD and $k$ is an optical wavenumber in vacuum. The SP mode is exited by the pumped QD. The enhancemnt of SP oscillations is inhibited by losses at the NP. The balance of these processes results in undamped stationary oscillations of the spaser dipole moment in absence of incident electromagnetic field (spasing).[8]

Though a spaser is a quantum device that requires a quantum-mechanical description (see, e.g., Refs. 1, 9-12), it could be described within the framework of classical electrodynamics since the near-field interaction determining the spaser operation is of classical nature. In the literature, the quantum-mechanical description is reduced to the modified Maxwell-Bloch equations[13] in which quantum-mechanical operators are substituted by $c$-numbers.[9, 14] The same assumptions are made for the description of a gain medium in terms of permittivity with negative losses.[15] The next step is to describe the QD as a particle made of an amplifying medium, which has permittivity with negative losses. This approach is utilized in Refs. 16-20, in which authors describe gain medium using a *linear* permittivity with a negative imaginary part. Although such an approach correctly demonstrates the lasing threshold of a nanolaser, it fails to reproduce nonlinear features of spasers, such as a stationary state with spasing, the spaser's behavior when pumping is above threshold, and the change in the population inversion by an external electromagnetic field. Certainly, the authors of the cited articles realized the necessity of taking nonlinear effects into account but attempts of doing so have a form of qualitative evaluations.[16, 21]

In this paper, we suggest an exactly solvable quasi-classical model of a spaser. The model is governed by the equations of classical electrodynamics and reflects the main features of spaser physics including the threshold transition to spasing as a Hopf bifurcation and predicts existing of the region where a spaser may be synchronized by external wave, the so-called Arnold tongue. We study two modifications of the model which differ by design but give qualitatively similar results.

## II. THE MODEL WITH SEPARATED NP AND QD

A classical system, which imitates the structure studied in Ref. 1, consists of a plasmonic NP (further PNP) with radius $r_P$ and a NP made of a gain medium (GNP) with radius $r_G$, as shown in Fig. 1. The NPs are separated by the distance $L$ much smaller than an optical wavelength in free space. The GNP has emission line at the frequency $\omega_0$.

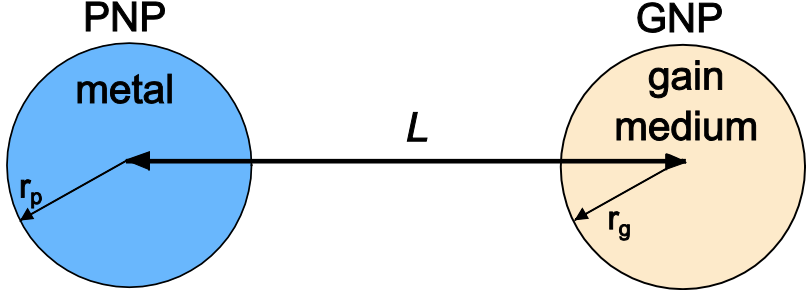

FIG. 1. A schematic drawing of the first model of spaser.

A gain medium is often described by an effective permittivity. The simplest expression for the permittivity suitable for such a description may be deduced from the Maxwell-Bloch equations,[13] which are commonly used in semiclassical description of lasers[22] and spasers.[9] In the framework of this approach, the evolution of electric field $\mathbf{E}$ is related to the macroscopic polarization $\mathbf{P}$ of a gain subsystem via classical Maxwell's equation:

$$\nabla \times \nabla \times \mathbf{E} + \frac{\varepsilon_0}{c^2}\frac{\partial^2 \mathbf{E}}{\partial t^2} = -\frac{4\pi}{c^2}\frac{\partial^2 \mathbf{P}}{\partial t^2}, \tag{1}$$

where $\varepsilon_0$ is the dielectric constant of the host matrix. The gain atoms embedded into the host medium are modeled as two-level systems with transition dipole moment $\boldsymbol{\mu}$ spread in the host matrix. Dynamics of the polarization and the population inversion $n$ is governed by the equations following from the density matrix formalism:[22]

$$\frac{\partial^2 \mathbf{P}}{\partial t^2} + \frac{2}{\tau_p}\frac{\partial \mathbf{P}}{\partial t} + \omega_0^2 \mathbf{P} = -\frac{2\omega_0 |\mu|^2}{\hbar} n\mathbf{E}, \tag{2a}$$

$$\frac{\partial n}{\partial t} + \frac{1}{\tau_n}(n - n_0) = \frac{1}{\hbar \omega_0}\mathrm{Re}\left(\mathbf{E}^* \cdot \frac{\partial \mathbf{P}}{\partial t}\right), \tag{2b}$$

where $\tau_p$ and $\tau_n$ are relaxation times for polarization and inversion, respectively, and $n_0$ stands for pumping of active atoms. Implying harmonic time dependence of the electric field and the polarization and excluding population inversion from this system we obtain the relation between the polarization $P$ and the electric field $E$ inside the medium, resulting in the following expression for nonlinear permittivity of a gain medium with an anti-Lorentzian profile:[15, 23]

$$\varepsilon_{gain}(\omega) = \varepsilon_0 + D_0 \frac{\omega_0}{\omega} \frac{-i + \frac{\omega^2 - \omega_0^2}{2\omega\Gamma}}{1 + \beta|E(\omega)|^2 + \left(\frac{\omega^2 - \omega_0^2}{2\omega\Gamma}\right)^2}, \qquad (3)$$

where $D_0 = 4\pi\mu^2 \tau_p n_0 / \hbar$ describes the population inversion, $D_0 = n_e - n_g$, $n_e$ and $n_g$ are populations of excited and ground states of the active atoms, respectively, $\beta = \mu^2 \tau_n \tau_p / \hbar^2$ and $\Gamma = 1/\tau_p$.

In the current study, we focus on the interaction between lossy and gain media and their scattering properties rather than on the description of an amplifying medium. For this reason, in the following calculations we take some realistic values of $\varepsilon_0$, $\Gamma$ and $D_0$ which are specified later. Note that zero value of $D_0$ does not provide gain. Negative population inversion corresponds to a lossy material. The electric field is measured in the units of $\beta^{-1/2}$, so the specific value of $\beta$ is of no importance. For metal permittivity we use the Drude formula $\varepsilon(\omega) = \varepsilon_\infty - \omega_{pl}^2 / \omega / (\omega + i\gamma)$. In order to fit actual experimental data,[24] we use the parameters corresponding to silver: $\varepsilon_\infty = 4.9$, $\omega_{pl} = 9.5\,\text{eV}$, and $\gamma = 0.05\,\text{eV}$.

Limiting ourselves to the dipole-dipole interaction of NPs, we consider the fields inside the NPs to be homogenous. The same approximation is made in the quantum mechanical consideration of spasers.[9, 11, 12]

Our goal is to find a nonzero solution $\mathbf{E}_{in}$ for the electric field inside the GNP in the absence of external (incident) field. The dipole moment of the GNP, $\mathbf{d}_G$, is related to $\mathbf{E}_{in}$ in the usual manner through its polarization $\mathbf{P}_G$:

$$\mathbf{d}_G = \frac{4\pi r_G^3}{3}\mathbf{P}_G = r_G^3 \frac{\varepsilon_{gain} - 1}{3}\mathbf{E}_{in}, \qquad (4a)$$

The dipole moment of the PNP induced by the dipole moment of the GNP is

$$\mathbf{d}_P = \alpha \mathbf{E}_G = \alpha \frac{3(\mathbf{d}_G \cdot \mathbf{n})\mathbf{n} - \mathbf{d}_G}{L^3}, \tag{4b}$$

where $\mathbf{E}_G$ is the elctric field of the GNP and $\alpha$ is the dipole polarizability of the PNP.[25] In particular cases of transverse or longitudinal polarizations, Eq. (4b) simplifies to the scalar expression $d_P = \alpha \kappa d_G / L^3$, where $\kappa$ is a geometrical factor depending on the polarization.[25] For the transverse polarizatrion, for which dipole moments are perpendicular to the line connetcing the centers of the NPs, $\kappa = -1$, and for the longitudinal polarization, $\kappa = 2$. Finally, we need the equation of continuity for the normal component of the electrical displacement on the surface of the GNP:

$$\varepsilon_{gain} \mathbf{E}_{in} \cdot \mathbf{n} = \mathbf{E}_{ext} \cdot \mathbf{n} = \left( \kappa \mathbf{d}_P / L^3 + 2\mathbf{d}_G / r_G^3 \right) \cdot \mathbf{n}, \tag{4c}$$

where $\mathbf{E}_{ext} = (\kappa \mathbf{d}_P)/L^3 + (3(\mathbf{d}_G \cdot \mathbf{n})\mathbf{n} - \mathbf{d}_G)/r_G^3$ is the electric field on the external surface of the GNP.

Assuming that $E_{in} \neq 0$, Eqs. (4) can be reduced to a single nonlinear equation determining $E_{in}$. Indeed, after substituting Eq. (4a) into Eq. (4b) and then substituting the resulted expression in Eq. (4c), one arrives at an equation containing $\omega$, $E_{in}$ and $D_0$:

$$\varepsilon_{gain}(\omega, E_{in}, D_0) = \frac{\kappa^2 \alpha(\omega) r_G^3 / L^6 + 2}{\kappa^2 \alpha(\omega) r_G^3 / L^6 - 1}, \tag{5}$$

or using Eq. (1) we obtain

$$\varepsilon_0 + D_0 \frac{\omega_0}{\omega} \frac{-i + \dfrac{\omega^2 - \omega_0^2}{2\omega \Gamma}}{1 + \beta |E_{in}|^2 + \left( \dfrac{\omega^2 - \omega_0^2}{2\omega \Gamma} \right)^2} = F(\omega), \tag{6}$$

where we introduce the notation

$$F(\omega) = \frac{\left( \kappa^2 \alpha(\omega) r_G^3 + 2L^6 \right)}{\left( \kappa^2 \alpha(\omega) r_G^3 - L^6 \right)} = \frac{\left( \left| \kappa^2 \alpha(\omega) r_G^3 \right|^2 + 2 + \kappa^2 r_G^3 \operatorname{Re} \alpha(\omega) L^6 \right)}{\left( \left| \kappa^2 \alpha(\omega) r_G^3 \right|^2 + L^{12} \right)} + i \frac{-\kappa^2 r_G^3 \operatorname{Im} \alpha(\omega) L^6}{\left( \left| \kappa^2 \alpha(\omega) r_G^3 \right|^2 + L^{12} \right)}.$$

The Eq. (6) determines the condition of existence of non-zero dipole moments $\mathbf{d}_P$ and $\mathbf{d}_G$ at zero incident field. Thus, it may be considered as an equation determining the condition of spasing for the particular model.

The left hand side of Eq. (6) depends on the frequency $\omega$, on the absolute value of the field inside the gain core $E_{in}$, and on the gain $D_0$, while the right hand side $F(\omega)$ depends on the frequency only. Since $\text{Im}\left[\kappa^2 \alpha r_G^3 / L^6\right] > 0$, one can see that $\text{Im}\left[F(\omega)\right] < 0$. For any positive $D_0$, the imaginary part of $\varepsilon_{gain}$ is negative as well, but is equal to zero for $D_0 = 0$. Thus, $\text{Im}\left[F(\omega)\right] < \text{Im}\, \varepsilon_{gain}$ for small values of gain, so that Eq. (6) does not have solutions. The minimal value of the gain $D_0$ for which Eq. (6) is satisfied can be considered as threshold gain for spasing, $D_{th}$. The corresponding frequency, at which Eq. (6) holds, is the spasing frequency, $\omega_{sp}$. Note that for $D_0 = D_{th}$ the dipole moments of NPs are equal to zero and $E_{in} = 0$.

An increase of $D_0 > D_{th}$ leads to nonzero dipole moments of NPs. Indeed, having set frequency to the value $\omega_{sp}$, one can recast Eq. (6) as

$$Z_{left}\left(D_0, E_{in}, \omega_{sp}\right) \equiv \frac{D_0}{1 + \beta \left|E_{in}\right|^2 + \left(\frac{\omega_{sp}^2 - \omega_0^2}{2\omega_{sp}\Gamma}\right)^2} = \frac{F(\omega_{sp}) - \varepsilon_0}{\left(-i + \frac{\omega_{sp}^2 - \omega_0^2}{2\omega_{sp}\Gamma}\right)} \frac{\omega_{sp}}{\omega_0} \equiv Z_{right}(\omega_{sp}), \quad (7)$$

where the right hand side of the equation $Z_{right}$ does not depend on gain $D_0$ and the internal field $E_{in}$. In fact, Eq. (7) determines the relation between the internal field $E_{in}$ and $D_0$. In the previous paragraph we have shown that $E_{in} = 0$ and $D_0 = D_{th}$ satisfy Eq. (7). An increase of $D_0$ accompanied with an increase of $E_{in}$ does not change the right hand side $Z_{right}$ because it depends neither on gain $D_0$ nor on internal field $E_{in}$. Thus,:

$$\frac{D_0}{1 + \beta \left|E_{in}(D_0)\right|^2 + \left(\frac{\omega_{sp}^2 - \omega_0^2}{2\omega_{sp}\Gamma}\right)^2} = \frac{D_{th}}{1 + 0 + \left(\frac{\omega_{sp}^2 - \omega_0^2}{2\omega_{sp}\Gamma}\right)^2}. \quad (8)$$

Hence, $E_{in}$, as well as the dipole moments of the NPs (Eqs. (4a) and (4b)), depends on $\sqrt{D_0 - D_{th}}$:

$$E_{in} = \sqrt{\frac{(D_0 - D_{th})}{\beta D_{th}}} \sqrt{1 + \left(\frac{\omega_{sp}^2 - \omega_0^2}{2\omega_{sp}\Gamma}\right)^2}, \tag{9a}$$

$$d_G = r_G^3 \frac{(\varepsilon_{gain} - 1)}{3} \sqrt{\frac{(D_0 - D_{th})}{\beta D_{th}}} \sqrt{\left[1 + \left(\frac{\omega_{sp}^2 - \omega_0^2}{2\omega_{sp}\Gamma}\right)^2\right]}, \tag{9b}$$

$$d_P = \alpha\kappa \frac{r_G^3}{L^3} \frac{(\varepsilon_{gain} - 1)}{3} \sqrt{\frac{(D_0 - D_{th})}{\beta D_{th}}} \sqrt{\left[1 + \left(\frac{\omega_{sp}^2 - \omega_0^2}{2\omega_{sp}\Gamma}\right)^2\right]}. \tag{9c}$$

This is similar to the Hopf bifurcation in regular lasers.[27] Indeed, for $D_0 = D_{th}$ the stable point $D = D_0$, $d_P = d_G = 0$ becomes unstable and a new stable point $D = D_{th}$ arises with square-root dependence of $d_G$ and $d_P$ on $D_0 - D_{th}$.

The spasing frequency, $\omega_{sp}$, can be found by using the following algorithm. Calculating real and imaginary parts of Eq. (6) at the threshold with $E_{in} = 0$, we obtain:

$$\varepsilon_0 + D_{th} \frac{\omega_0}{\omega_{sp}} \frac{(\omega_{sp}^2 - \omega_0^2)/2\omega_{sp}\Gamma}{1 + \left((\omega_{sp}^2 - \omega_0^2)/2\omega_{sp}\Gamma\right)^2} = \text{Re}[F(\omega_{sp})], \tag{10a}$$

$$D_{th} \frac{\omega_0}{\omega_{sp}} \frac{-1}{1 + \left((\omega_{sp}^2 - \omega_0^2)/2\omega_{sp}\Gamma\right)^2} = \text{Im}[F(\omega_{sp})] \tag{10b}$$

.Excluding the unknown $D_{th}$ from this system we arrive at the transcendental equation which determines the spasing frequency:

$$\varepsilon_0 = \text{Re}[F(\omega_{sp})] + \frac{\omega_{sp}^2 - \omega_0^2}{2\omega_{sp}\Gamma} \text{Im}[F(\omega_{sp})]. \tag{11}$$

Once $\omega_{sp}$ is found numerically or analytically, one obtains threshold pumping from Eq. (10):

$$D_{th} = -\frac{\omega_{sp}}{\omega_0}\left(1 + \left((\omega_{sp}^2 - \omega_0^2)/2\omega_{sp}\Gamma\right)^2\right)\text{Im}[F(\omega_{sp})]. \tag{12}$$

## III. THE EXACTLY SOLVABLE CORE-SHELL MODEL

The assumption of the field uniformity inside the NPs made in the previous section is not realistic. Correct consideration requires finding a solution of a nonlinear Laplace problem, which would to very cumbersom calculations but would not change results qualitatively. Fortunatelly, there is a special geometry that permits to solve the nonlinear problem analyticaly and to obtain an exact solution. Being inspired by theoretical consideration[17, 19] and successful manufacturing[29] of a composite core-shell nanoparticles we take such a structure as a model for a spaser. Indeed, the uniform field inside the core is a solution to the nonlinear Laplace equation. The strength of this field may be considered as the eigenvalue of the problem.

Below we consider a gain spherical core coated with a metallic plasmonic shell (Fig. 2) and find its response on the external harmonic field.

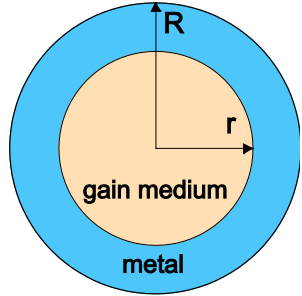

FIG. 2. A schematic drawing of a core-shell spaser.

The spasing state of such a system can be found by imposing boundary conditions at two interfaces. For simplicity, we fix the homogenous electric field inside the gain core $\mathbf{E}_{core}(\mathbf{r}) \equiv \mathbf{E}$ and write the field in a metallic shell as a linear response in the form

$$\mathbf{E}_{shell}(\mathbf{r}) = b\mathbf{E} - c\mathbf{E}/r^3 + 3c(\mathbf{E}\cdot\mathbf{n})\mathbf{n}/r^3, \qquad (13)$$

and the field outside the nanoparticle as

$$\mathbf{E}_{out}(\mathbf{r}) = a\mathbf{E}/r^3 + 3a(\mathbf{E}\cdot\mathbf{n})\mathbf{n}/r^3, \qquad (14)$$

which vanishes at the infinity. In Eqs. (13) and (14) and below $\mathbf{n} = \mathbf{r}/r$. The three unknown coefficients, $a$, $b$, and $c$, are found from the boundary conditions for the electromagnetic field at the inner and outter surface of the metallic shell:

$$\begin{aligned}
\mathbf{D}_{core} \cdot \mathbf{n}\big|_r &= \mathbf{D}_{shell} \cdot \mathbf{n}\big|_r \\
\mathbf{D}_{shell} \cdot \mathbf{n}\big|_R &= \mathbf{D}_{out} \cdot \mathbf{n}\big|_R \\
\mathbf{E}_{core} \times \mathbf{n}\big|_r &= \mathbf{E}_{shell} \times \mathbf{n}\big|_r \\
\mathbf{E}_{shell} \times \mathbf{n}\big|_R &= \mathbf{E}_{out} \times \mathbf{n}\big|_R
\end{aligned} \qquad (15)$$

Substituting Eqs. (13) and (14) in the boundary conditions (15), we arrive at the following system of equations:

$$\begin{cases} \varepsilon_{gain}(|E|) = \varepsilon_{shell}(b + 2c/r^3) \\ 1 = b - c/r^3 \\ b - c/R^3 = -a/R^3 \\ \varepsilon_{shell}(b + 2c/R^3) = 2a/R^3 \end{cases}. \qquad (16)$$

Again, system (16) can be recasted in the form of a single equation representing the condition of spasing:

$$\varepsilon_{gain}(\omega, E, D_0) = \varepsilon_{shell}\left(1 - 3\frac{(2 + \varepsilon_{shell})R^3}{2(\varepsilon_{shell} - 1)r^3 + (2 + \varepsilon_{shell})R^3}\right). \qquad (17)$$

Similarly to Eq. (5), the right hand side of Eq. (17) depends on frequency $\omega$ only, while its left hand side is permittivity of the gain medium which depends on $E$ and $D_0$. Thus, the core-shell system also demostrates the Hopf bifurcation and $\sqrt{D_0 - D_{th}}$-dependence of the spaser dipole moment above threshold. Indeed, using the Eq. (1) we obtain formula which coincides with Eq. (9a), so that the internal field and the dipole moment of the core-shell spaser are:

$$E_{in} = \sqrt{\frac{(D_0 - D_{th})}{\beta D_{th}}} \sqrt{1 + \left(\frac{\omega_{sp}^2 - \omega_0^2}{2\omega_{sp}\Gamma}\right)^2}, \qquad (18a)$$

$$d_{core-shell} = aE_{in}, \qquad (18b)$$

where $a$ is determined from system (16):

$$a = \left[-(1 + 2\varepsilon_{shell})(\varepsilon_{shell} - \varepsilon_{gain})r^3 + (\varepsilon_{shell} - 1)(2\varepsilon_{shell} + \varepsilon_{gain})R^3\right](9\varepsilon_{shell})^{-1}. \qquad (19)$$

## IV. LOSS COMPENSATION IN THE MODEL WITH SEPARATED NP AND QD

Including a gain medium into a metamaterial made of plasmonic NPs turns the metamaterial into a matrix filled with spasers. These spasers can be used for loss compensation in the system.[10, 30-34] One can expect that the wave propagation may be discribed in terms of the effective permittivity. Such a discription implies that the spaser should respond to the external field at least linearly and should oscilate with the frequency of the driving field. When the loss compenssation occurs, the imaginary part of the spaser dipole moment is equal to zero.

In Refs. 34-36 the possibility of loss compensation was studied by computer simulation. In order to be able to use the effective permittivity to describe the spaser response, the authors considered very short pulses of the external wave. It was assumed that during the pulse the population inversion does not change. Thus, despite of using nonlinear equations, the authors obtain results of the linear theory.

The present toy model of the spaser allows us to consider nonlinear response of the spaser. Above pumping threshold, a spaser is a self-oscillating system with the fixed frequency and the amplitude. Therefore, in this regime spasers are not very convenient for loss compensation for a wide range of frequencies.[37, 38] Even though such a spaser can be synchronized by external optical field, so that it oscillates with the fequency of that field[38] and losses can be compensated at certain frequencies and amplitudes of that external field,[31] the amplitude of spaser dipole oscillations weakly depends on the amplitude of the external field. The value of this amplitude is about the same as the amplitude of oscillations of a non-driven spaser.

The response of a spaser operating below pumping threshold is more suitable for compensating losses in a metamaterial matrix. Indeed, below pumping threshold a spaser does not oscillate without an external field. Such a spaser is always synchronized by the external field. The amplitude of the dipole oscillations nonlinearly depends on the strength of the external field. The question remains whether the driven below-threshold spaser could actually compensate losses. Here we answer this question within the framework of the model of separated NPs.

In order to calculate the response of a driven below-threshold spaser on the external incident field, we should include a respective term $E_{ext}$ into the system of equations (4) for either transverse or logitudinal polarization. Assuming that the dipole moments of both NPs oscillate with the frequency of the external field $\omega$, we can write:

$$d_G = r_G^3 \frac{\varepsilon_{gain} - 1}{3} E_{in}, \tag{20a}$$

$$d_P = \alpha \left( E_{ext} + \kappa d_G / L^3 \right), \tag{20b}$$

$$\varepsilon_{gain} E_{in} = E_{ext} + \kappa d_P / L^3 + 2 d_G / r_G^3. \tag{20c}$$

The solution of Eqs. (20) gives values of dipole moments of NPs for a given frequency and the amplitude of the external field. Since the separation distance $L$ is of the order of the subwavelength, the spaser radiates in the far zone as a single electric dipole having dipole moment $d_G + d_P$. In Fig. 3(a) we plot the total dipole moment of the driven below-threshold spaser for the case of longitudinal poalrization ($\kappa = 2$). One can see that there is a regime in which the imaginary part of the total dipole moment field vanishes, i.e. losses in NPs are compensated by gain below spasing threshold. Loss compensation is achieved in the regime of above-threshold pumping as well (Fig. 3b), however, as we discuss above, this regime is less suitable for this purpose.

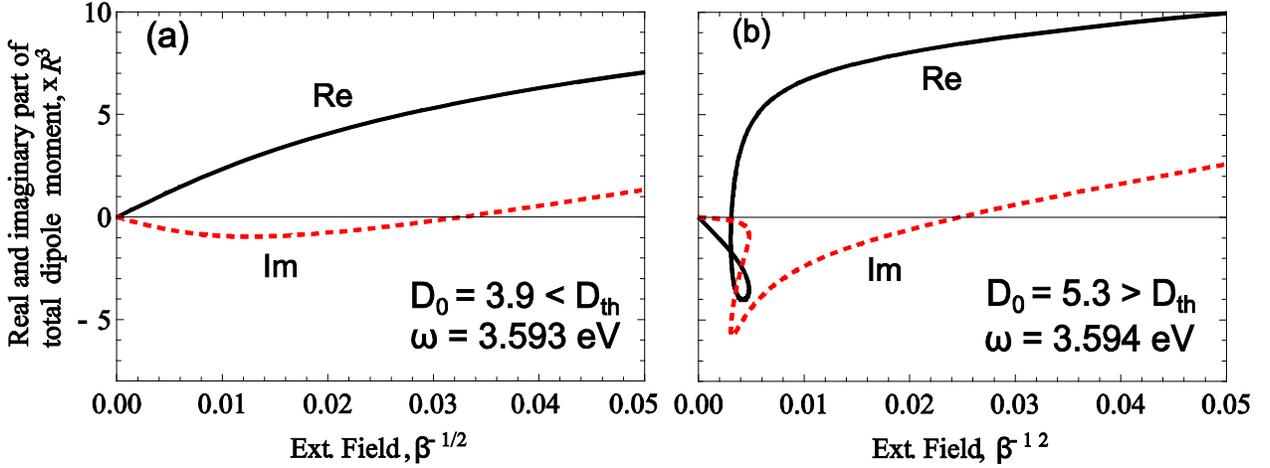

FIG. 3. Response of spaser on the external oscillating field in (a) below- and (b) above- threshold regimes with longitudinal polarization. In both figures the following parameters are used: $r_P = r_G = 20 \text{ nm}$, $L = 2.5 r_P$, $\omega_0 = 3.6 \text{ eV}$, $\varepsilon_0 = 2$, and $\Gamma = 0.01 \text{ eV}$. Solid and dashed lines show real and imaginary parts of the dipole moment, respectively.

## V. SPASER SYNCRONIZATION AND THE ARNOLD TONGUE

Fig. 3(b) features a typical response of a driven above-threshold spaser in the model of separated GNP and PNP. One can see that there is a region in which three different steady state solutions correspond to a given frequency and amplitude of the external field. However, only one of them is stable. To shed light on this issue we consider equations describing the temporal evolution of the dipole moment of a spaser driven by the external oscillating field. The analysis is done for a core-shell spaser, however, the main findings hold for the model of separated NPs as well.

To investigate the time evolution of the stationary state we cannot consider the external wave as a plane wave with a constant amplitude, but should consider a slowly varying long pulse of the external field $\mathbf{E}_{ext}(t) = \mathbf{E}_{slow}(t)e^{-i\Omega t}$ and corresponding dipole moment, $\mathbf{d}(t) = \mathbf{d}_{slow}(t)e^{-i\Omega t}$, induced in a core-shell spaser (see for details Ref. 40). Here $\mathbf{d}_{slow}(t)$ and $\mathbf{E}_{slow}(t)$ are slowly varying envelopes, which Fourier transformations include only frequencies that are much smaller than the central frequency $\Omega$. The external field and dipole moment should be related via the nonlinear operator,

$$\hat{\alpha}^{-1}(t, E_{in}(t))\mathbf{d}(t) = \mathbf{E}_{ext}(t). \tag{21}$$

The explicit form of the operator $\hat{\alpha}^{-1}(E_{in})$ is not known because it depends on the field inside the GNP, $E_{in}(t)$, which is a long pulse as well, whereas the permittivity (1) is written for a harmonic field with the amplitude $E(\omega)$. To make Eqs. (21) and (1) consistent with each other, we define the operator $\hat{\alpha}^{-1}(E_{in})$ through its action on the harmonic field as follows:

$$\alpha^{-1}(\omega, E_{in}(\omega))\mathbf{d}(\omega) = \mathbf{E}_{ext}(\omega), \tag{22}$$

where $E_{in}(\omega)$ is the value that should be put into Eq. (1).

Applying the Fourier transformation to Eq. (21) we get:

$$\mathbf{E}_{ext}(t) = \int \alpha^{-1}(\omega, E_{in})\mathbf{d}(\omega)e^{-i\omega t}d\omega = e^{-i\Omega t}\int \alpha^{-1}(\Omega+\nu, E_{in})\mathbf{d}_{slow}(\nu)e^{-i\nu t}d\nu. \tag{23}$$

Since the main contribution into the integral (23) is given by harmonics having $v \ll \Omega$, we obtain

$$\mathbf{E}_{ext}(t) \approx e^{-i\Omega t} \int \left[ \alpha^{-1}(\Omega, E_{in}) + v \frac{d\alpha^{-1}(\Omega, E_{in})}{d\Omega} \right] \mathbf{d}_{slow}(v) e^{-ivt} dv =$$
$$= e^{-i\Omega t} \alpha^{-1}(\Omega, E_{in}) \int \mathbf{d}_{slow}(v) e^{-ivt} dv + e^{-i\omega_0 t} \frac{d\alpha^{-1}(\Omega, E_{in})}{d\Omega} \int v \mathbf{d}_{slow}(v) e^{-ivt} dv \quad (24)$$
$$= e^{-i\Omega t} \left[ \alpha^{-1}(\Omega, E_{in}) \mathbf{d}_{slow}(t) + i \frac{d\alpha^{-1}(\Omega, E_{in})}{d\Omega} \frac{d\mathbf{d}_{slow}(t)}{dt} \right],$$

where the term with $d/dt$ accounts for small broadening of dipole moment spectra (see also Refs. 41, 42). Cancelling the oscillating factor $e^{-i\Omega t}$ at both sides of Eq. (24) we arrive at the desired equation describing temporal evolution of the spaser dipole moment in slowly varying external field with the central frequency $\Omega$:

$$i \frac{d\alpha^{-1}}{d\Omega} \frac{d}{dt} \mathbf{d}_{slow}(t) + \alpha^{-1} \mathbf{d}_{slow}(t) = \mathbf{E}_{slow}(t). \quad (25)$$

Provided $\mathbf{E}_{slow}(t) \equiv \mathbf{E}$ is constant, Eq. (25) has a stationary solution $\mathbf{d}_{slow} = \alpha \mathbf{E}$. In order to study how small perturbations of this solution evolve with time, let us consider a perturbation in the form $\mathbf{d}_{slow} = \alpha \mathbf{E} + \delta \mathbf{d} e^{\Lambda t}$. The instability growth rate $\Lambda$ is then given by

$$\Lambda = i\alpha^{-1}(\Omega, E_{in}) \left( \frac{d\alpha^{-1}(\Omega, E_{in})}{d\Omega} \right)^{-1}. \quad (26)$$

The stationary solution of Eq. (25) becomes unstable when $\operatorname{Re} \Lambda > 0$. In Fig. 4 we plot the real part of the instability growth rate for both below- and above-threshold regimes of a driven core-shell spaser.

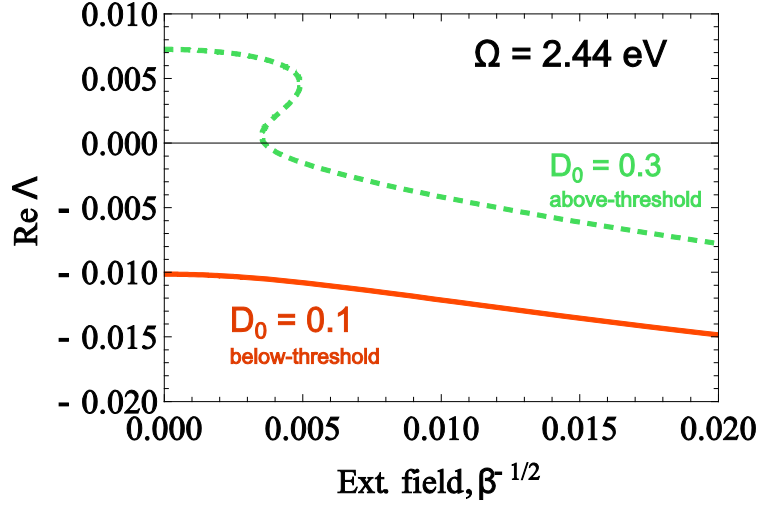

FIG. 4. Real parts of the instability growth rate $\Lambda$ for below- (solid line) and above-threshold (dashed line) regimes of a driven spaser. The parameters of spaser are: $r = 15\,\text{nm}$, $R = 20\,\text{nm}$, $\varepsilon_0 = 4$, $\Gamma = 0.05\,\text{eV}$, $\omega_0 = 2.45\,\text{eV}$.

In the below-threshold regime of a driven spaser, $\text{Re}\,\Lambda < 0$. Thus, the steady-state oscillations are stable with respect to small perturbations. When pumping increases, so that $D_0$ exceeds $D_{th}$, a region in which $\text{Re}\,\Lambda > 0$ arises. In this region, the steady-state solution becomes unstable. In the above-threshold regime without external field, the spaser oscillates with its spasing frequency (see Eq. (11)). When the external field is applied, depending on the field amplitude and frequency detuning, the spaser may or may not oscillate with the frequency of this field. When it has the same frequency as the external field, the steady-state solution is stable and spaser is synchronized with the external field. The region in which the synchronization takes place is known the Arnold tongue.[43] This region is shown in Fig. 5 for $D_0 = 0.3$. When the frequency of the external field is tuned to the spasing frequency, $\omega_{sp}$, the synchronization occurs for vanishingly small amplitude of the external field.

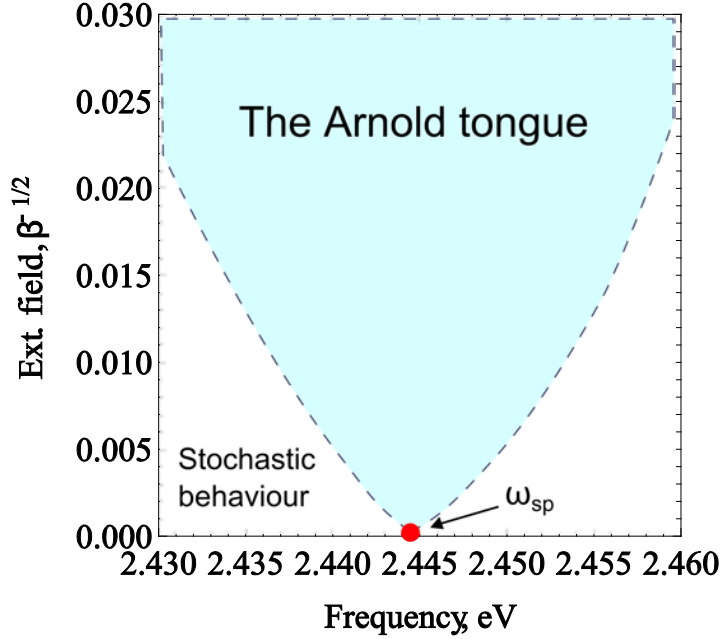

FIG 5. The region corresponding to the Arnold tongue in which synchronization of a spaser by an external field occurs for gain coefficient $D_0 = 0.3$.

## VI. DISCUSSION

The model presented here reproduces the general features of the quantum description of a spaser including the pumping threshold for spasing, the Hopf bifurcation, and the exsitence a region of spaser sinchronization with an external field – the Arnold tongue. Our semiclassical model also predicts the possibility of loss compensation by a spaser operating below threshold (more rigorous quantum mechanical consideration of this problem will be published elsewhere).

Our model also reveals inconsistencies in linear models of nanolaser. In particular, the authors of Ref. 17 consider a metal-coated nanolaser and report a lasing turn-off above the threshold. This result is in disagreement with the experimental observation of spasing in core-shell nanolasers and with general laser theory.[27] The authors make a suggestion that the on/off behavior of lasing in coated nanoparticles is caused by detuning of the resonance when the gain is added. However, in our study we show that the spasing frequency does not depend on gain and is a function only of the nanolaser geometry (see Eq. (11)).

Using our model, it is interesting to look at the discussion concerning the possibility of loss compensation in plasmonic systems with gain (see Ref. 38 and comments to this work). In Ref. 38, Stockman argues that in a resonant plasmonic structure, Ohmic losses are compensated for by gain when spasing occurs. Indeed, this argument is valid for a closed system in which there is no incoming and outgoing radiation. In this case, loss compensation and lasing simply coincide.

In an open system coupled with the radiation, it is necessary to compensate for both Ohmic and radiation losses for spasing to occur. In this case, as we show above for a spaser below-threshold, lossless scattering of an incoming wave may occur when the system does not spase. This happens because the magnitude of dipole oscillations is smaller than in the above-threshold spaser and the pumping energy is therefore sufficient to compensate for the loss. This situation is analogous to the scheme suggested in Ref. 30, in which Ohmic losses in the illuminated photonic crystal composed of alternating metallic and dielectric amplifying layers are compensated below the lasing threshold.

## ACKNOWLEDGEMENTS

The authors are indebted to R. E. Noskov for helpful discussion of relation (25). This work was supported by RFBR Grants Nos. 10-02-91750, 11-02-92475 and 12-02-01093 and by a PSC-CUNY grant.

[8] M. I. Stockman, Opt. Express **19**, 22029 (2011).

[9] I. E. Protsenko, A. V. Uskov, O. A. Zaimidoroga, V. N. Samoilov, and E. P. O'Reilly, Phys. Rev. A **71**, 063812 (2005).

[10] A. K. Sarychev and G. Tartakovsky, Phys. Rev. B **75**, 085436 (2007).

[11] A. S. Rosenthal and T. Ghannam, Phys. Rev. A **79**, 043824 (2009).

[12] E. S. Andrianov, A. A. Pukhov, A. V. Dorofeenko, A. P. Vinogradov, and A. A. Lisyansky, Phys. Rev. B **85**, 035405 (2012).

[13] M. O. Scully and M. S. Zubairy, *Quantum Optics* (Cambridge University Press, Cambridge, 1997).

[14] A. N. Lagarkov, A. K. Sarychev, V. N. Kissel, and G. Tartakovsky, Phys. Usp. **52**, 959 (2009).

[15] A. A. Zyablovsky, A. V. Dorofeenko, A. A. Pukhov, and A. P. Vinogradov, J. Commun. Techn. Electr. **56**, 1139 (2011).

[16] N. M. Lawandy, App. Phys. Lett. **85**, 5040 (2004).

[17] J. A. Gordon and R. W. Ziolkowski, Opt. Express **15**, 2622 (2007).

[18] A. Mizrahi, V. Lomakin, B. A. Slutsky, M. P. Nezhad, L. Feng, and Y. Fainman, Opt. Lett. **33**, 1261 (2008).

[19] X. F. Li and S. F. Yu, Opt. Lett. **35**, 2535 (2010).

[20] M. I. Stockman, Phil. Trans. R. Soc. A **369**, 3510 (2011).

[21] A. Veltri and A. Aradian, Phys. Rev. B **85**, 115429 (2012).

[22] R. H. Pantell and H. E. Puthoff, *Fundamentals of quantum electronics* (Wiley, New York, 1969).

[23] S. Solimeno, B. Crosignani, and P. Di Porto, *Guiding, Diffraction, and Confinement of Optical Radiation* (Academic Press, Orlando, 1986).

[24] P. B. Johnson and R. W. Christy, Phys. Rev. B **6**, 4370 (1972).

[25] C. F. Bohren and D. R. Huffman, *Absorption and Scattering of Light by Small Particles* (Wiley, New York, 1983).